\documentclass[conference, a4paper]{IEEEtran}

\usepackage{cite}
\usepackage{amsmath}
\usepackage{amssymb}
\usepackage{amsfonts}
\usepackage{graphicx}
\usepackage[dvipsnames]{xcolor}
\usepackage[a4paper, total={184mm,239mm}]{geometry}
\usepackage[final]{microtype}
\usepackage{import}
\usepackage[free-standing-units,per-mode=repeated-symbol,mode=text,binary-units=true,detect-weight=true,detect-family=true]{siunitx}
\usepackage{glossaries}
\usepackage{cleveref}
\usepackage{pgfplots}
\usepackage{tikz}
\usepackage{booktabs}
\usepackage{textcomp}
\usepackage{enumitem}
\usepackage{url}
\usepackage{nth}

\usetikzlibrary{scopes, calc, shapes, arrows, positioning}
\tikzset{>=latex}

\pgfplotsset{compat=1.13}
\pgfplotsset{width=\linewidth, height=7cm}
\pgfplotsset{every x tick label/.append style={font=\small}}
\pgfplotsset{every y tick label/.append style={font=\small}}

\makeatletter
\let\MYcaption\@makecaption
\makeatother
\usepackage[font=footnotesize]{subcaption}
\makeatletter
\let\@makecaption\MYcaption
\makeatother

\DeclareSIUnit\core{core}
\DeclareSIUnit\tile{tile}
\DeclareSIUnit\request{request}
\DeclareSIUnit\cycle{cycle}
\DeclareSIUnit\erlang{E}
\DeclareSIUnit\flop{FLOP}
\DeclareSIUnit\flops{FLOPS}
\DeclareSIUnit\gate{GE}
\DeclareSIUnit\op{OP}
\DeclareSIUnit\ops{OPS}
\DeclareSIUnit\bps{bps}
\DeclareSIUnit\Bps{Bps}
\DeclareSIUnit\ipc{IPC}

\newacronym[longplural={Scratchpad Memories}]{SPM}{SPM}{Scratchpad Memory}
\newacronym{ACE}{ACE}{AXI Coherent Extensions}
\newacronym{AMBA}{AMBA}{Advanced Microcontroller Bus Architecture}
\newacronym{APB}{APB}{Advanced Peripheral Bus}
\newacronym{API}{API}{Application Programming Interface}
\newacronym{ASIC}{ASIC}{Application-Specific Integrated Circuit}
\newacronym{AVX}{AVX}{Advanced Vector Extension}
\newacronym{AXI}{AXI}{Advanced eXtensible Interface}
\newacronym{BLAS}{BLAS}{Basic Linear Algebra Subprograms}
\newacronym{CHI}{CHI}{Coherent Hub Interface}
\newacronym{CMOS}{CMOS}{Complementary Metal-Oxide-Semiconductor}
\newacronym{CNN}{CNN}{Convolutional Neural Network}
\newacronym{CPU}{CPU}{Central Processing Unit}
\newacronym{CSR}{CSR}{Control and State Register}
\newacronym{CTS}{CTS}{Clock Tree Synthesis}
\newacronym{DLP}{DLP}{Data Level Parallelism}
\newacronym{DMA}{DMA}{Direct Memory Access}
\newacronym{DRAM}{DRAM}{Dynamic Random-Access Memory}
\newacronym{DSP}{DSP}{Digital Signal Processing}
\newacronym{DUT}{DUT}{Device Under Test}
\newacronym{ECL}{ECL}{Emitter-Coupled Logic}
\newacronym{FBB}{FBB}{Forward Body-Biasing}
\newacronym{FDSOI}{FD-SOI}{Fully Depleted Silicon on Insulator}
\newacronym{FMA}{FMA}{Fused Multiply-Add}
\newacronym{FPGA}{FPGA}{Field-Programmable Gate Array}
\newacronym{FPU}{FPU}{Floating Point Unit}
\newacronym{GPGPU}{GPGPU}{General-Purpose \acrlong{GPU}}
\newacronym{GPU}{GPU}{Graphics Processing Unit}
\newacronym{HDL}{HDL}{Hardware Description Language}
\newacronym{HERO}{HERO}{Heterogeneous Embedded Research Platform}
\newacronym{HPC}{HPC}{High-Performance Computing}
\newacronym{ILP}{ILP}{Instruction Level Parallelism}
\newacronym{IOT}{IoT}{Internet-of-Things}
\newacronym{IPC}{IPC}{Instructions Per Cycle}
\newacronym{IPU}{IPU}{Image Processing Unit}
\newacronym{ISA}{ISA}{Instruction Set Architecture}
\newacronym{LSU}{LSU}{Load/Store Unit}
\newacronym{LVT}{LVT}{low voltage threshold}
\newacronym{MIMD}{MIMD}{multiple instruction, multiple data}
\newacronym{MMU}{MMU}{Memory Management Unit}
\newacronym{MUL}{MUL}{multiplier}
\newacronym{MVL}{MVL}{maximum vector length}
\newacronym{NUMA}{NUMA}{non-uniform memory access}
\newacronym{NOC}{NoC}{Network-on-Chip}
\newacronym{PCIe}{PCIe}{Peripheral Component Interconnect Express}
\newacronym{PC}{PC}{Program Counter}
\newacronym{PE}{PE}{processing element}
\newacronym{PL}{PL}{Programmable Logic}
\newacronym{PMCA}{PMCA}{Programmable Manycore Accelerator}
\newacronym{PSL}{PSL}{Power Service Layer}
\newacronym{PTE}{PTE}{page-table entry}
\newacronym{PTW}{PTW}{page-table walker}
\newacronym{PULP}{PULP}{Parallel Ultra Low Power}
\newacronym{RAW}{RAW}{read-after-write}
\newacronym{RBB}{RBB}{Reverse Body-Biasing}
\newacronym{ROB}{ROB}{Reorder Buffer}
\newacronym{RTL}{RTL}{Register Transfer Level}
\newacronym{RVT}{RVT}{Regular Voltage Threshold}
\newacronym{RoCC}{RoCC}{Rocket Custom Coprocessor Interface}
\newacronym{SCM}{SCM}{Storage Class Memory}
\newacronym{SIMD}{SIMD}{single instruction, multiple data}
\newacronym{SIMT}{SIMT}{single instruction, multiple thread}
\newacronym{SLDU}{SLDU}{Slide Unit}
\newacronym{SLVT}{SLVT}{super-low voltage threshold}
\newacronym{SM}{SM}{Streaming Multiprocessor}
\newacronym{SRAM}{SRAM}{Static Random-Access Memory}
\newacronym{SSE}{SSE}{Streaming SIMD Extension}
\newacronym{SVE}{SVE}{Scalable Vector Extension}
\newacronym{TLP}{TLP}{Thread Level Parallelism}
\newacronym{TxnID}{TxnID}{Transaction ID}
\newacronym{VAC}{VAC}{Vector Access}
\newacronym{VC}{VC}{virtual channel}
\newacronym{VCONV}{VCONV}{Vector Conversion}
\newacronym{VEX}{VEX}{Vector Execute}
\newacronym{VFU}{VFU}{vector functional unit}
\newacronym{VID}{VID}{Vector Instruction Decode}
\newacronym{VIS}{VISSUE}{Vector Instruction Issue}
\newacronym{VLIW}{VLIW}{Very Long Instruction Word}
\newacronym{VLOOP}{VLOOP}{Vector Loop}
\newacronym{VLR}{VLR}{vector length register}
\newacronym{VLSU}{VLSU}{Vector Load/Store Unit}
\newacronym{VNB}{VNB}{Von Neumann Bottleneck}
\newacronym{VRF}{VRF}{Vector Register File}
\newacronym{VT}{VT}{vector thread}
\newacronym{WAR}{WAR}{write-after-read}
\newacronym{WAW}{WAW}{write-after-write}
\newacronym{DCT}{DCT}{discrete cosine transform}

\definecolor{color1}{HTML}{256DFF}
\definecolor{color2}{HTML}{45CCB0}
\definecolor{color3}{HTML}{9775CA}
\definecolor{color4}{HTML}{C83737}
\colorlet{colorAlert}{Red}

\newcommand\eg{e.g.,\ }
\newcommand\ie{i.e.,\ }

\newcommand\mempool{MemPool}
\newcommand\topology[1]{Top\ensuremath{_{\text{#1}}}}

\begin{document}

\title{\mempool{}: A Shared-L1 Memory Many-Core Cluster with a Low-Latency Interconnect}
\author{
  \IEEEauthorblockN{Matheus Cavalcante}
  \IEEEauthorblockA{ETH Zürich\\
    Zürich, Switzerland\\
    matheusd \emph{at} iis.ee.ethz.ch}\and
  \IEEEauthorblockN{Samuel Riedel}
  \IEEEauthorblockA{ETH Zürich\\
    Zürich, Switzerland\\
    sriedel \emph{at} iis.ee.ethz.ch}\and
  \IEEEauthorblockN{Antonio Pullini}
  \IEEEauthorblockA{GreenWaves Technologies\\
    Grenoble, France\\
    pullinia \emph{at} iis.ee.ethz.ch}\and
  \IEEEauthorblockN{Luca Benini}
  \IEEEauthorblockA{ETH Zürich\\
    Zürich, Switzerland\\
    Università di Bologna\\
    Bologna, Italy\\
    lbenini \emph{at} iis.ee.ethz.ch}}
\maketitle

\begin{abstract}
  A key challenge in scaling shared-L1 multi-core clusters towards many-core (more than 16 cores) configurations is to ensure low-latency and efficient access to the L1 memory.
  In this work we demonstrate that it is possible to scale up the shared-L1 architecture: We present \mempool{}, a \SI{32}{\bit} many-core system with \num{256} fast RV32IMA ``Snitch'' cores featuring application-tunable execution units, running at \SI{700}{\mega\hertz} in typical conditions (TT/\SI{0.80}{\volt}\kern-.1em/\SI{25}{\celsius}).
  \mempool{} is easy to program, with all the cores sharing a global view of a large L1 scratchpad memory pool, accessible within at most \num{5} cycles.
  In \mempool{}'s physical-aware design, we emphasized the exploration, design, and optimization of the low-latency processor-to-L1-memory interconnect.
  We compare three candidate topologies, analyzing them in terms of latency, throughput, and back-end feasibility.
  The chosen topology keeps the average latency at fewer than \num{6} cycles, even for a heavy injected load of \SI{0.33}{\request\per\core\per\cycle}.
  We also propose a lightweight addressing scheme that maps each core private data to a memory bank accessible within one cycle, which leads to performance gains of up to \SI{20}{\percent} in real-world signal processing benchmarks.
  The addressing scheme is also highly efficient in terms of energy consumption since requests to local banks consume only half of the energy required to access remote banks. Our design achieves competitive performance with respect to an ideal, non-implementable full-crossbar baseline.
\end{abstract}



\section{Introduction}
\label{sec:introduction}

The failure of Dennard scaling~\cite{Dennard1974} has implied a power wall for computing, limiting processor frequencies~\cite{Esmaeilzadeh2011}. To achieve high performance under a limited power budget, core count scaling has been used instead. Multi-core architectures are the norm today, allowing for high performance and efficient computing on a wide range of applications.

There are many flavors of multi-core architectures.
Some consist of a few general-purpose high-performance cores sharing a large cache, such as Arm's Cortex-A77~\cite{ArmCortexA772019} and Intel's Core-i9 processors~\cite{IntelCorei92019}.
Others are highly specialized processor arrays, usually with a specialized interconnection network adapted to the intended application domain~\cite{Blake2009}, such as Google's Pixel Visual Core~\cite{Redgrave2018}, an \gls{IPU} with a 2D mesh network of \num{256} computing units that communicate with near-neighbor fixed-size messages.

We focus on a common architectural pattern for building multi-core architectures, namely a cluster of simple cores sharing L1 memory through a low-latency interconnect~\cite{Ayed2016}.
We can find instances of this architectural pattern across many different domains, from the streaming processors of \glspl{GPU}~\cite{Lindholm2008}, to the ultra-low-power domain with GreenWaves' GAP8 processor~\cite{GreenWavesGAP82019}, to high-performance embedded signal processing with the Kalray processor clusters~\cite{Dinechin2013}, to aerospace applications with the Ramon Chips' RC64 system~\cite{Ginosar2016}.
However, as we will detail in \Cref{sec:related-work}, these clusters only scale to low tens of cores and suffer from long memory access latency due to the topology of their interconnects.

In this paper, we set out to design and optimize the first scaled-up \emph{many-core} system with shared low-latency L1 memory.
To this end, we propose \mempool{}, a \SI{32}{\bit} RISC-V-based system, with \num{256} small cores sharing a large pool of \gls{SPM}.
In the absence of contention, all the \gls{SPM} banks are accessible within \num{5} cycles.
The contributions of this paper are:
\begin{itemize}
\item The physical-aware design of \mempool{}'s architecture, with particular emphasis on the exploration, design, and optimization of a low-latency processor-to-L1-memory interconnection network (\Cref{sec:architecture});
\item The creation of a lightweight and transparent memory addressing scheme that keeps the memory region most often accessed by a core---\eg the stack---in a memory bank close by, with minimal access latency (\Cref{sec:scrambling-logic});
\item The complete physical implementation and performance, power, and area analysis of a large cluster in an advanced \textsc{GlobalFoundries} 22FDX \gls{FDSOI} technology node (\Cref{sec:performance-analysis,sec:phys-impl}).
\end{itemize}

\mempool{} runs at \SI{700}{\mega\hertz} in typical conditions (\SI{480}{\mega\hertz} in worst-case conditions).
The critical path of the design is dominated by wire delay (\SI{37}{\percent}), with \num{27} out of its \num{36} gates being either buffers or inverter pairs.
Its processor-to-L1 interconnect has an average latency of fewer than \num{6} cycles, even for a heavy injected load of \SI{0.33}{\request\per\core\per\cycle}.
Our addressing scheme helps to keep the memory requests in local banks accessible within one cycle, which leads to performance gains up to \SI{20}{\percent} in real-world benchmarks.
This scheme is also highly effective in terms of energy consumption since local memory requests consume only half of the energy required for remote memory accesses. In a nutshell, we demonstrate in this paper that we can scale the core count of an L1-shared cluster to ten times more cores than what was previously considered achievable, with cycle counts on various benchmarks that are comparable with an ideal, non-implementable full-crossbar baseline.


\section{Related Work}
\label{sec:related-work}

In this work, we focus on architectures that feature a cluster of simple cores sharing low-latency L1 memory~\cite{Ayed2016}, which is a very common architectural pattern.
The latest Nvidia Ampere \glspl{GPU}, for example, have ``streaming multiprocessors'' with \num{32} double-precision floating-point cores each, sharing \SI{192}{\kibi\byte} of L1 memory~\cite{NvidiaA1002020}.
Similar architectural patterns can also be found in the embedded and ultra-low-power domain.
GreenWaves' GAP8~\cite{GreenWavesGAP82019} is an \gls{IOT} application-class processor with eight cores sharing \SI{128}{\kibi\byte} of L1 memory.
The Snitch cluster~\cite{Zaruba2020} features a cluster of eight RV32IMA cores sharing \SI{128}{\kibi\byte} of private \gls{SPM} accessible within \num{2} cycles.

It is commonly believed that the number of cores in a single L1-shared cluster is bound to the low-tens limit. To scale the core count of many-core systems into the hundreds, memory sharing is usually done at some high-latency hierarchy level.
Moving to multiple clusters creates challenges in terms of programmability~\cite{Dinechin2013}:
tile-based systems are usually connected by meshes that have long access latency and usually require \gls{NUMA} models of computation.

An example of multi-cluster design is Kalray's MPPA-256. It integrates \num{256} user cores into \num{16} clusters.
Each MPPA-256 cluster has its own private address space~\cite{Dinechin2013}, with the clusters connected by specialized \glspl{NOC}.
Memory sharing is done at the level of main memory.
This design achieves high efficiency by compromising on memory sharing between clusters, thus circumventing the need for low-level cache coherence protocols.
Tilera's TILE-Gx series, on the other hand, equips each tile processor with an L2 cache, and inter-tile and main memory communication are provided by five 2D mesh networks~\cite{Wentzlaff2007}.
Ramon Chips' RC64 system brings this design pattern to the harsh aerospace conditions~\cite{Ginosar2016}.
Its \num{64} cores have private \glspl{SPM}, and share access to \SI{4}{\mebi\byte} of on-chip memory, accessible through a logarithmic network.

Clearly, there is a push toward high core-count many-core architectures to tackle embarrassingly parallel problems, such as image processing and machine learning.
Google's Pixel Visual Core~\cite{Redgrave2018}, for example, is an \gls{IPU} with specialized stencil processors interconnected with a ring network.
Within each stencil processor, an array of \num{256} lanes communicate through a rigid read-neighbor network.
This design is highly specialized for systolic algorithms, which can take advantage of data sharing between neighbors.
The rigidity of memory allocation and the long access latency reduce the applicability of this highly efficient design to other non-systolic algorithms.

The main novelty of our work is to demonstrate we can scale up the shared-L1 processor cluster in a region that was considered to be completely infeasible by all previous related works, namely hundreds of cores, with extremely competitive performance and efficiency.


\section{Architecture}
\label{sec:architecture}

\mempool{} has a large multi-banked pool of L1 memory, shared among all the cores.
A fully connected non-blocking logarithmic interconnect between the \num{256} cores would be infeasible due to routing congestion and area scaling~\cite{Michelogiannakis2009}.
In this work, we show this can be overcome through a hierarchical approach at the topological and physical level, placing the cores and memory banks in a regular array connected by low-diameter networks.
The following sections go bottom-up over the main hierarchical modules that compose \mempool{}.

\subsection{Interconnect architecture}
\label{sec:network-architecture}

\mempool{} has two parallel interconnects, one routing the cores' requests to the \gls{SPM} banks, and one routing read responses back.
The basic element of both interconnects is a single-stage $m \times n$ crossbar switch, connecting $m$ masters to $n$ slaves.
An optional elastic buffer can be inserted at each output of the switch, after address decoding and round-robin arbitration,
to break any combinational paths crossing the switch~\cite{Dimitrakopoulos2014}.

\begin{figure}[h]
  \centering
  \includegraphics[width=0.70\linewidth]{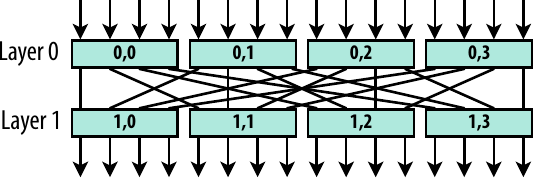}
  \caption{Topology of a \num{16x16} radix-\num{4} butterfly.
    The boxes represent the base routing element, a \num{4x4} logarithmic crossbar switch.}
  \label{fig:bfl}
\end{figure}

With the access latency constraint in mind, the interconnect building blocks are kept as streamlined as possible.
They do not provide transaction ordering, with this task offloaded to the cores.
Moreover, since they are used to transmit single-word-requests from the cores, the network does not use virtual channels.
We use oblivious routing since there is a single path for each master/slave pair.
In terms of network topology, two-dimensional mesh networks were discarded due to their bad latency scaling.
We also avoided Ogras and Marculescu's approach~\cite{Ogras2006}, which optimizes the latency of a mesh network by iteratively adding long-ranging links to the topology, but requires an advanced routing algorithm.
\mempool{} uses a combination of fully-connected crossbars and minimal radix-\num{4} butterfly networks, whose topology can be seen in \Cref{fig:bfl}.

\subsection{Tile}
\label{sec:tile}

\begin{figure}[h]
  \centering
  \includegraphics[width=0.95\linewidth]{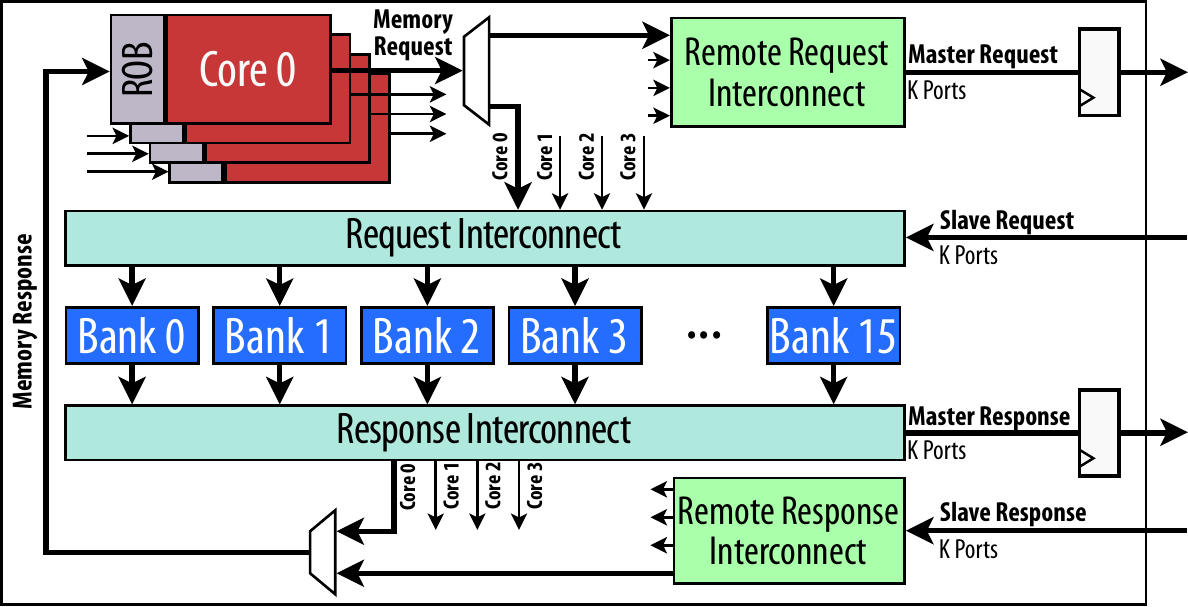}
  \caption{Architecture of \mempool{} tiles, with $K$ request ports and $K$ response ports to remote tiles.}
  \label{fig:arch_tile}
\end{figure}

At the base of \mempool{}'s hierarchy, we have tiles, whose architecture can be seen in \Cref{fig:arch_tile}.
The tiles are based on Snitch, a \SI{21}{\kilo\gate} single-issue single-stage RISC-V-based RV32IMA core~\cite{Zaruba2020}, whose small area allows for massive replication.
Snitch supports a configurable number of outstanding load instructions, which is useful to hide the \gls{SPM} access latency.

The tile contains \num{16} memory banks, and each core has a dedicated port to access them with one cycle latency.
The cores share $K$ master ports to access remote tiles.
An address decoder at the output of the cores statically decides where to send the cores' requests.
Each tile has $K$ slave request ports, receiving memory requests from remote tiles.
There is a register boundary at the master request and response ports.
Both request and response interconnects are realized as fully-connected crossbars.
Requests hold metadata to route them back to the correct core and ensure their proper ordering by the \gls{ROB}.

Inside each tile, we have a \num{4}-way L1 instruction cache.
The cache has a \SI{32}{\bit} \gls{AXI} refill port.
These ports can be connected to a low-overhead refill-network (\eg a ring), which is noncritical, and hence it is not further discussed in this work.

\subsection{\mempool{} cluster}
\label{sec:cluster-1}

We evaluated three network topologies for the global interconnection network between tiles.

\subsubsection{\topology{1} -- single \num{64x64} radix-\num{4} butterfly}
\label{sec:single-num64x64-num4}

In this configuration, each tile has a single remote port for communication with remote tiles, \ie $K=1$.
A single \num{64x64} radix-\num{4} butterfly network, with a single pipeline stage midway through its $\log_4(64) = 3$ layers, connects the tiles.
Therefore, data in any remote memory bank can be accessed within \num{5} cycles.
Inside the tile, two \num{4x1} crossbars arbitrate the core requests and the memory responses.
This design creates a bottleneck at the tile boundary, since the traffic of \num{4} cores is concentrated through a single port.

\subsubsection{\topology{4} -- four parallel \num{64x64} radix-\num{4} butterflies}
\label{sec:topol-four-parall}

To reduce the traffic bottleneck at the tile boundary, we evaluated a system that replicates the global \num{64x64} butterfly interconnect four times.
That is, each tile has four master request and response ports, each associated with their own \num{64x64} interconnect.
Each master request port is dedicated to a core, \ie the remote request interconnect is effectively a point-to-point connection.

\subsubsection{\topology{H} -- hierarchical approach}
\label{sec:topol-hier-appr}

\begin{figure}[h]
  \centering

  \begin{minipage}[t]{0.65\linewidth}
    \centering
    \includegraphics[width=.9\linewidth]{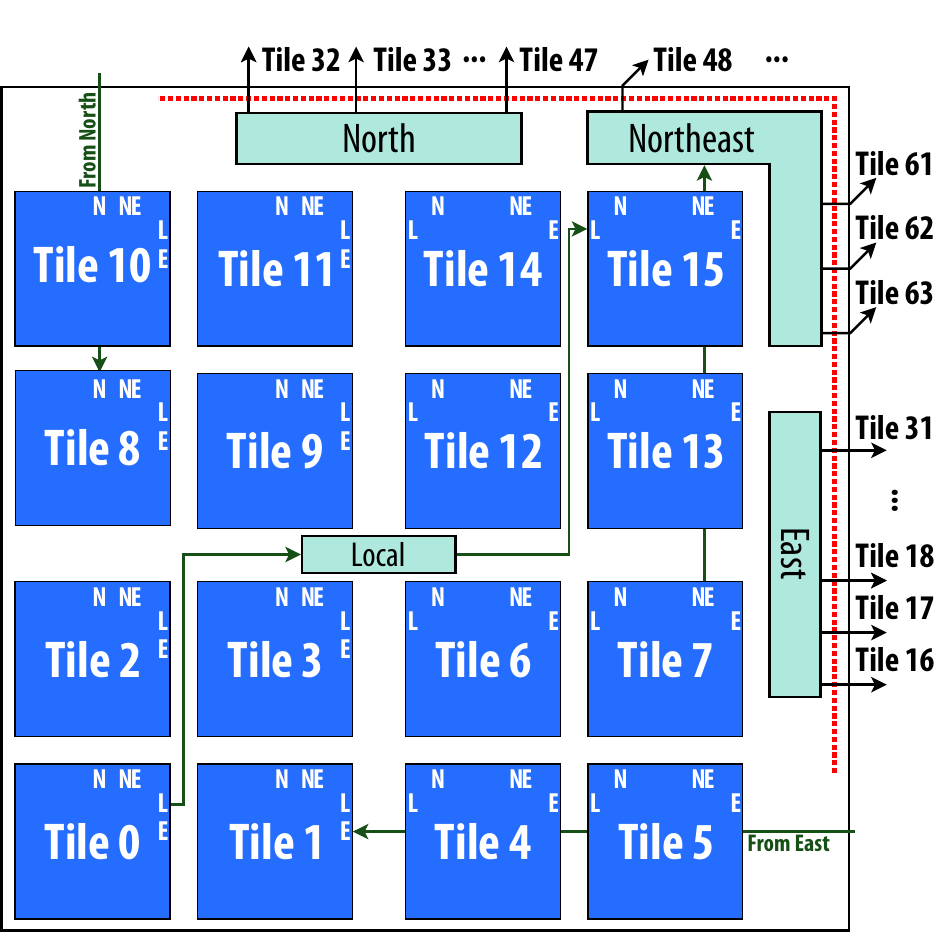}
    \subcaption{Local group.}
    \label{fig:arch_group}
  \end{minipage}\hfill%
  \begin{minipage}[t]{0.30\linewidth}
    \centering
    \includegraphics[width=\linewidth]{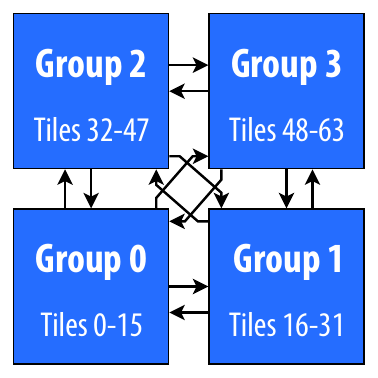}
    \subcaption{Cluster.}
    \label{fig:arch_cluster}
  \end{minipage}

  \caption{\mempool{}'s \topology{H} architecture. Dashed lines indicate a register boundary.}
  \label{fig:arch_mempool}
\end{figure}

Both \topology{1} and \topology{4} have a uniform access pattern, with a \num{5}~cycle access latency between any two tiles.
Hence, requests between two tiles need to cross the whole interconnect, regardless of how physically close they are.
This leads to decreased efficiency and increased routing congestion due to longer paths towards the center of the design.
We introduce a new level of hierarchy to maintain the bandwidth of \topology{4}, while avoiding long detours between neighboring tiles.
\Cref{fig:arch_group} shows a \emph{local group} of \num{16} tiles.
Cores access remote memory banks in the same local group within \num{3} cycles thanks to the local interconnect, a \num{16x16} fully-connected crossbar.

The \mempool{} cluster is composed of four local groups, as shown in \Cref{fig:arch_cluster}.
Inside each local group, the \emph{north} (N), \emph{northeast} (NE), and \emph{east} (E) \num{16x16} radix-\num{4} butterfly networks are responsible for communication between local groups.
Each tile has corresponding N, NE, and E ports, and a \emph{local} (L) port to access tiles within the same local group.
A \num{4x4} crossbar inside each tile routes the requests to the correct port.
There is a register boundary at the local groups' master interfaces, increasing the zero-load access latency of a memory bank in a remote local group to \num{5} cycles.


\section{Hybrid Addressing Scheme}
\label{sec:scrambling-logic}

\begin{figure}[bth]
  \centering
  \includegraphics[width=0.9\linewidth]{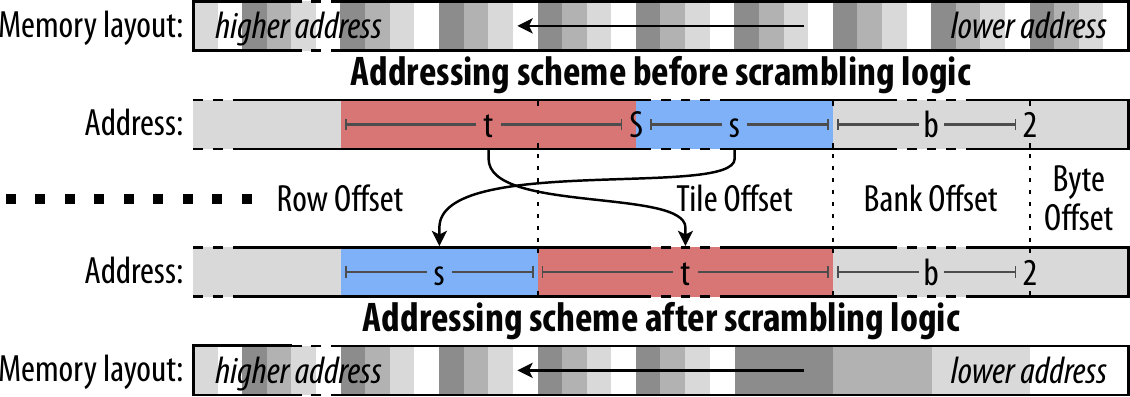}
  \caption{Hybrid addressing scheme via the scrambling logic.
    The upper and lower parts show the fully interleaved and the hybrid memory map, respectively.
    The outer bars visualize the memory map, with the shades representing different tiles to which the addresses are mapped.
    The inner bars are the addresses, with the scrambling in between leading from one scheme to the other.}
  \label{fig:scrambling_logic}
\end{figure}

\mempool{} has a sequentially interleaved memory mapping across all memory banks in order to minimize banking conflicts.
However, this also implies that most memory requests target remote tiles.
Optimally, all cores' requests would remain in the local tile, which would lower the latency and power consumption.
With the scrambling logic, visualized in \Cref{fig:scrambling_logic}, we transform an interleaved memory map into a \emph{hybrid} one, by adding \emph{sequential regions} in which contiguous addresses target a single tile.
The top half shows the classical fully interleaved memory scheme.
The address and memory map at the bottom is a hybrid memory map created by swapping the address bits.

With an interleaved memory addressing scheme, the addresses are interpreted as follows.
The first two bits are the byte offset, after which $b$ bits identify one of the $2^b$ banks of each tile.
The next $t$ bits distinguish between the $2^t$ tiles.
The remaining address bits are interpreted as the row offset within each bank.

Consider each tile with a sequential memory region of $2^S$ bytes, or $2^s$ rows in the tile's banks.
Since the banks inside the same tile are still accessed interleaved, we leave the byte and bank offsets untouched.
The next $s$ bits represent part of the tile offset, but we need them to represent the banks' next row within the same tile.
Therefore, we shift them $t$ bits to the left---where the row offset starts---and fill them with the $t$ bits we replaced.
This creates $2^{t}$ sequential regions, one for each tile.
In total, we dedicate the first $2^{S+t}$ bytes to sequential regions.
We leave the subsequent bytes interleaved by conditionally applying the scrambling to addresses inside the sequential memory region.


The hybrid addressing scheme's key benefit is giving the programmer the additional capability to store private data, such as a core's stack, in the same tile.
It reduces the number of transactions between tiles,
making better use of the tiles' fully-connected, high-throughput crossbar.
Sequential memory regions are prone to banking conflicts. However, by only mapping private data to the sequential region, the cores' accesses remain distributed across all banks.
In contrast to aliasing or completely private memories, we do not complicate programmability but, by applying the same address transformation for all cores, give all cores the same memory view and keep the L1 memory region contiguous and shared.
The beneficiary of the sequential region are programs that make heavy use of the stack or work mainly on local data.
The scrambling logic can be efficiently implemented in hardware with a wire crossing and a multiplexer.


\section{Performance analysis}
\label{sec:performance-analysis}

\subsection{Interconnect architecture}
\label{sec:interc-arch}

In this section, we analyze the three proposed network topologies in terms of average latency and throughput, as a function of the injected load $\lambda$ (measured in requests per core per cycle).
The results were extracted using an extensive cycle-accurate \gls{RTL} simulation.
Each core is replaced by a synthetic traffic generator, which generates new requests following a Poisson process of rate $\lambda$.
The requests have a random uniformly distributed destination memory bank.

\begin{figure}[ht]
  \centering
  \begin{minipage}[ht]{0.5\linewidth}
    \resizebox{\linewidth}{!}{
    \begin{tikzpicture}[/tikz/font=\footnotesize]
      \begin{axis}[
        xlabel = {Injected load (\si{\request\per\core\per\cycle})},
        xmin = 0,
        xmax = 0.5,
        xtick distance = 0.1,
        ylabel = {Throughput (\si{\request\per\core\per\cycle})},
        ymin = 0,
        ymax = 0.5,
        ytick distance = 0.1,
        height = 5cm,
        grid   = major,
        legend pos = north west,
        legend style = {font=\footnotesize}]

        \addplot [color1, very thick] table [x expr = \thisrowno{0}*0.001] {fig/interco/result_thru_top1};
        \addlegendentry{\topology{1}};

        \addplot [color2, very thick] table [x expr = \thisrowno{0}*0.001] {fig/interco/result_thru};
        \addlegendentry{\topology{4}};

        \addplot [color3, very thick] table [x expr = \thisrowno{0}*0.001] {fig/interco/result_thru_toph};
        \addlegendentry{\topology{H}};
      \end{axis}
    \end{tikzpicture}}
    \subcaption{Throughput.}
    \label{fig:thru}
  \end{minipage}\hfill%
  \begin{minipage}[ht]{0.5\linewidth}
    \resizebox{\linewidth}{!}{
    \begin{tikzpicture}[/tikz/font=\footnotesize]
      \begin{axis}[
        xlabel = {Injected load (\si{\request\per\core\per\cycle})},
        xmin = 0,
        xmax = 0.5,
        xtick distance = 0.1,
        ylabel = {Average latency (\si{\cycle})},
        ymin = 0,
        ymax = 20,
        restrict y to domain=*0:200,
        ytick distance = 4,
        height = 5cm,
        grid   = major]

        \addplot [color1, very thick] table [x expr = \thisrowno{0}*0.001] {fig/interco/result_avglat_top1};
        \addplot [color2, very thick] table [x expr = \thisrowno{0}*0.001] {fig/interco/result_avglat};
        \addplot [color3, very thick] table [x expr = \thisrowno{0}*0.001] {fig/interco/result_avglat_toph};
      \end{axis}
    \end{tikzpicture}}
    \subcaption{Average latency.}
    \label{fig:latency}
  \end{minipage}
  \caption{Network analysis of the three proposed network topologies, in terms of throughput and average round-trip latency, as a function of the injected load.}
  \label{fig:analysis}
\end{figure}
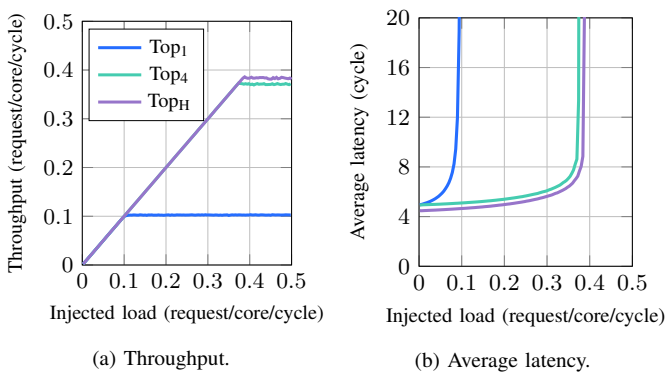

\Cref{fig:thru} shows the different topologies' throughput, with an increasing load.
At a load of \SI{0.10}{\request\per\core\per\cycle}, \topology{1} becomes congested, while \topology{4} and \topology{H} support almost four times that load, about \SI{0.38}{\request\per\core\per\cycle}.
\topology{H}'s throughput is slightly higher than \topology{4}'s, due to its smaller diameter.

As a counterpart to \Cref{fig:thru}, \Cref{fig:latency} shows the average round-trip latency of the requests for an increasing load.
It elevates the point where the topologies become congested by showing the explosion of the average latency.
The average latency of \topology{H} only reaches \num{6} cycles at a network load of \SI{0.33}{\request\per\core\per\cycle}.
Due to \topology{H}'s three-cycle latency to a local group, it achieves a smaller average latency than \topology{4}.
Both results imply that the \topology{1}'s traffic concentration at the tiles' ports leads to unacceptable performance degradation.

\subsection{Hybrid addressing scheme}
\label{sec:scrambling-logic-1}

To evaluate the performance impact of the hybrid addressing scheme, we analyze \topology{H} taking the hybrid addressing scheme into account.
The traffic generator creates uniformly-distributed requests to the local tile's sequential region with probability $p_{\text{local}}$, and outside of this region with probability $1-p_{\text{local}}$.

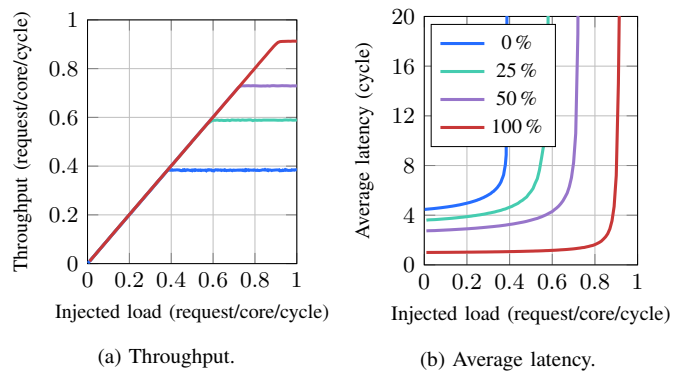
\begin{figure}[ht]
  \centering
  \begin{minipage}[ht]{0.5\linewidth}
    \resizebox{\linewidth}{!}{
    \begin{tikzpicture}[/tikz/font=\footnotesize]
      \begin{axis}[
        xlabel = {Injected load (\si{\request\per\core\per\cycle})},
        xmin = 0,
        xmax = 1,
        xtick distance = 0.2,
        ylabel = {Throughput (\si{\request\per\core\per\cycle})},
        ymin = 0,
        ymax = 1,
        restrict y to domain=*0:1,
        ytick distance = 0.2,
        height = 5cm,
        grid   = major]

        \addplot [color1, very thick] table [x expr = \thisrowno{0}*0.001] {fig/scrambling/throughput_0};
        \addplot [color2, very thick] table [x expr = \thisrowno{0}*0.001] {fig/scrambling/throughput_250};
        \addplot [color3, very thick] table [x expr = \thisrowno{0}*0.001] {fig/scrambling/throughput_500};
        \addplot [color4, very thick] table [x expr = \thisrowno{0}*0.001] {fig/scrambling/throughput_1000};
      \end{axis}
    \end{tikzpicture}}
    \subcaption{Throughput.}
    \label{fig:thru_scr}
  \end{minipage}\hfill%
  \begin{minipage}[ht]{0.5\linewidth}
    \resizebox{\linewidth}{!}{
    \begin{tikzpicture}[/tikz/font=\footnotesize]
      \begin{axis}[
        xlabel = {Injected load (\si{\request\per\core\per\cycle})},
        xmin = 0,
        xmax = 1,
        xtick distance = 0.2,
        ylabel = {Average latency (\si{\cycle})},
        ymin = 0,
        ymax = 20,
        restrict y to domain=*0:300,
        ytick distance = 4,
        height = 5cm,
        grid   = major,
        legend pos = north west,
        legend style = {font=\footnotesize}]]

        \addplot [color1, very thick] table [x expr = \thisrowno{0}*0.001] {fig/scrambling/latency_0};
        \addlegendentry{\SI{0}{\percent}}
        \addplot [color2, very thick] table [x expr = \thisrowno{0}*0.001] {fig/scrambling/latency_250};
        \addlegendentry{\SI{25}{\percent}}
        \addplot [color3, very thick] table [x expr = \thisrowno{0}*0.001] {fig/scrambling/latency_500};
        \addlegendentry{\SI{50}{\percent}}
        \addplot [color4, very thick] table [x expr = \thisrowno{0}*0.001] {fig/scrambling/latency_1000};
        \addlegendentry{\SI{100}{\percent}}
      \end{axis}
    \end{tikzpicture}}
    \subcaption{Average latency.}
    \label{fig:latency_scr}
  \end{minipage}
  \caption{Network analysis of \topology{H} with our hybrid addressing scheme, as a function of the injected load, for different probabilities of requesting data in the local tile's sequential region $p_{\text{local}}$.}
  \label{fig:thru_scramble}
\end{figure}

\Cref{fig:thru_scr} shows the throughput of \topology{H} for different $p_{\text{local}}$.
It shows a clear trend of an increased throughput for a larger $p_{\text{local}}$.
The scrambling logic, or using local memory in general, can vastly improve the system's throughput by preventing the congestion in the global interconnect, besides lowering the overall average access latency as seen in \Cref{fig:latency_scr}.
An application making \SI{25}{\percent} of its accesses to the stack, mapped at the sequential region, can gain up to \SI{50}{\percent} in performance by using the scrambling logic, without changing the code.

\subsection{Benchmarks}
\label{sec:benchmarks}

In this section, we benchmark \mempool{} with three real-world highly-parallelizable signal processing benchmarks' runtime:
\begin{description}[font=\normalfont\textit]
\item[matmul:] a matrix multiplication of two \num{64x64} matrices, for which accesses are predominantly remote;
\item[2dconv:] a 2D discrete convolution with a \num{3x3} kernel, for which all accesses are local, except for cores working on windows that require data from two tiles;
\item[dct:] a 2D \gls{DCT} operating on \num{8x8} blocks residing in local memory. It uses the stack to store intermediate results, \ie all accesses are local, given the stack is mapped to local banks.
\end{description}

We derive our baseline systems from a Snitch cluster~\cite{Zaruba2020}, which we ideally scale up to a cluster of \num{256} cores connected to \num{1024} banks through a fully-connected crossbar.
The systems assume an idealized interconnect with no routing conflicts that allows all banks to be accessed within one cycle. 
This idealized network is physically infeasible due to high routing congestion with reasonable clock rates.
We use two baseline systems, with (\topology{XS}) and without (\topology{X}) scrambling logic, to compare to the respective \topology{$\blacklozenge$S} and \topology{$\blacklozenge$} \mempool{} systems (where \topology{$\blacklozenge$} stands for any of the topologies defined in \Cref{sec:cluster-1}).

\begin{figure}[t]
  \centering
  \resizebox{\linewidth}{!}{
  \begin{tikzpicture}
    \begin{axis}[
      height=4.5cm,
      width=\linewidth,
      ybar=0pt,
      enlargelimits=0,
      scaled y ticks = false,
      bar width=8pt,
      legend image code/.code={
        \draw [#1] (0cm,-0.1cm) rectangle (0.4cm,0.2cm); },
      legend style={
        at={(0.5,-0.2)},
        anchor=north,
        transpose legend,
        legend columns=2,
        /tikz/every even column/.append style={column sep=0.3cm}
      },
      legend cell align={left},
      ylabel={Relative performance},
      ymin=0,
      enlarge y limits={value=0.05,upper},
      ymajorgrids=true,
      ytick={0,0.2,0.4,0.6,0.8,1},
      ytick style={draw=none},
      yticklabel style={/pgf/number format/.cd,fixed,fixed zerofill,precision=1},
      symbolic x coords={Start,\emph{matmul},\emph{2dconv},\emph{dct},End},
      xtick=data,
      xtick pos=left,
      xmin=\emph{matmul},
      xmax=\emph{dct},
      enlarge x limits=0.25,
      ]
      \pgfplotstableread{fig/benchmarks/result_bench_rel_arch}\loadedtable;
      \addplot [fill=color1!70!white] table[x=Kernels, y=Top1] {\loadedtable};
      \addplot [fill=color1!80!black] table[x=Kernels, y=Top1+S] {\loadedtable};
      \addplot [fill=color2         ] table[x=Kernels, y=Top4] {\loadedtable};
      \addplot [fill=color2!80!black] table[x=Kernels, y=Top4+S] {\loadedtable};
      \addplot [fill=color3         ] table[x=Kernels, y=TopH] {\loadedtable};
      \addplot [fill=color3!80!black] table[x=Kernels, y=TopH+S] {\loadedtable};
      \addplot [fill=color4         ] table[x=Kernels, y=TopX] {\loadedtable};
      \addplot [fill=color4!80!black] table[x=Kernels, y=TopX+S] {\loadedtable};
      \addplot[black, line width=0.4mm, dotted, sharp plot, update limits=false] coordinates {(Start,1) (End,1)};
      \legend{
        \topology{1},
        \topology{1S},
        \topology{4},
        \topology{4S},
        \topology{H},
        \topology{HS},
        \topology{X},
        \topology{XS}}
    \end{axis}
  \end{tikzpicture}}
  \caption{Performance of the three benchmarks on all topologies, relative to the baselines.
    \topology{$\blacklozenge$S} represents the \topology{$\blacklozenge$} with scrambling logic.}
  \label{fig:benchmark}
\end{figure}
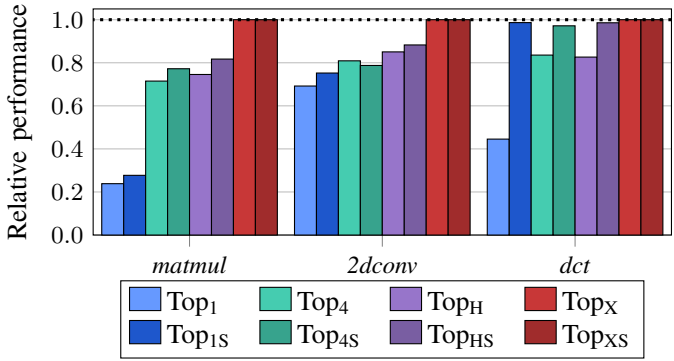

\Cref{fig:benchmark} shows the benchmarks' performance, normalized by the performance achieved on the baseline architectures.
\topology{H} generally beats \topology{4}, and they both outperform \topology{1} by a factor of three in the extreme cases.
A big performance difference is visible in \emph{matmul}, which has many remote accesses.
For \emph{dct}, the three topologies with the scrambling logic perform equally well, as they all operate on data mapped to local banks.
Without the scrambling logic, the stacks become spread over all tiles, leading to a significant performance penalty, especially for \topology{1}.

\topology{H} performs very close to the baseline for all benchmarks, with a performance penalty of at most \SI{20}{\percent} for \emph{matmul} due to its non-local access pattern.
The hybrid memory map increases the locality of the data accesses, improving overall performance.
With \emph{dct}, we match the baseline since we only do local accesses.

\section{Physical implementation}
\label{sec:phys-impl}

In this section, we analyze the feasibility of \mempool{} using the \topology{1}, \topology{4}, and \topology{H} topologies.
We also analyze them in terms of power, performance, and area results.

\subsection{Methodology}
\label{sec:methodology}

\mempool{} was synthesized for \textsc{GlobalFoundries} 22FDX \gls{FDSOI} technology using Synopsys Design Compiler Graphical 2019.12.
We used floorplanning information during synthesis to improve timing correlation with the back-end design.
Each tile has \SI{2}{\kibi\byte} of instruction cache, and \SI{16}{\kibi\byte} of \gls{SPM}---\ie the \mempool{} cluster has \SI{1}{\mebi\byte} of L1 \gls{SPM}.
The back-end flow was carried out with Synopsys IC Compiler~II 2019.12, targeting \SI{500}{\mega\hertz} at worst-case conditions (SS/\SI{0.72}{\volt}\kern-.1em/\SI{125}{\celsius}).
\mempool{}'s power results were extracted with switching activities obtained by simulating the benchmarks on a netlist back-annotated with post-place-and-route delay information.
We used Synopsys PrimeTime 2019.12 to carry out the sign-off timing extraction at worst-case conditions and power analysis at typical conditions (TT/\SI{0.80}{\volt}\kern-.1em/\SI{25}{\celsius}).

\subsection{Tile implementation}
\label{sec:tile-implementation}

\begin{figure}[h]
  \centering
  \resizebox{.45\linewidth}{!}{
  \begin{tikzpicture}[/tikz/font=\small]
    \node[anchor=south west,inner sep=0] (image) at (0,0) {\includegraphics[width=.50\linewidth]{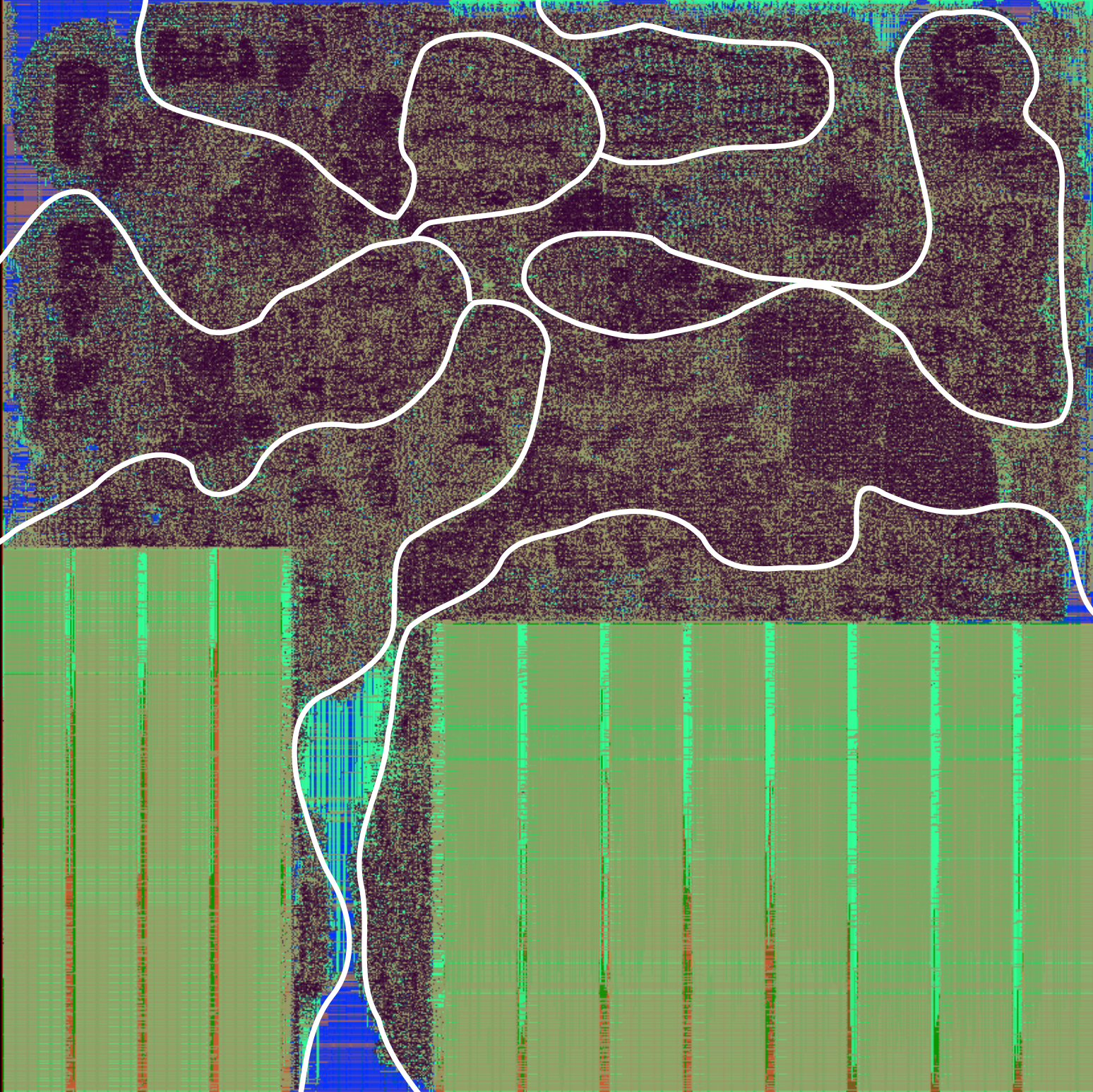}};
    \begin{scope}[x={(image.south east)}, y={(image.north west)}]
      \node [white] at (0.70,0.22) {\textbf{\gls{SPM}}};
      \node [white] at (0.14,0.25) {\textbf{I\$}};
      \node [white] at (0.15,0.65) {\textbf{C0}};
      \node [white] at (0.20,0.80) {\textbf{C1}};
      \node [white] at (0.30,0.90) {\textbf{C2}};
      \node [white] at (0.90,0.80) {\textbf{C3}};
      \node [white] at (0.90,0.80) {\textbf{C3}};
      \node [white] at (0.70,0.59) {\textbf{Interconnect}};
    \end{scope}
  \end{tikzpicture}}
  \caption{Placed and routed \topology{H} tile, as a \SI{425x425}{\micro\meter} macro.}
  \label{fig:fp_tile}
\end{figure}

Due to the size and the regularity of \mempool{}'s design, we used a hierarchical implementation flow.
The tile was implemented as a square \SI{425x425}{\micro\meter} macro (\SI{908}{\kilo\gate}).
The most complex tile---\ie \topology{H}'s tile---is shown in \Cref{fig:fp_tile}.

The tile's critical path (\num{53} gates long) starts at a register after the instruction cache, passing through the \nth{2} Snitch core and the request interconnect, before arriving at a \gls{SPM} bank.
The tile achieves a utilization of \SI{72.8}{\percent}, the area is dominated by the instruction cache (\SI{23.6}{\percent}) and by the L1 \gls{SPM} (\SI{40.2}{\percent}).

\subsection{\mempool{} cluster implementation}
\label{sec:memp-clust-impl}

The \mempool{} cluster is implemented as a \SI{4.6x4.6}{\milli\meter} macro, \ie \SI{55}{\percent} of the design area is covered by the tiles.
The area overhead was driven by congestion, which is the main constraint of the design, particularly at the center of the design.

\begin{figure}[h]
  \centering

  \begin{minipage}[t]{0.50\linewidth}
    \centering
    \includegraphics[width=.9\linewidth]{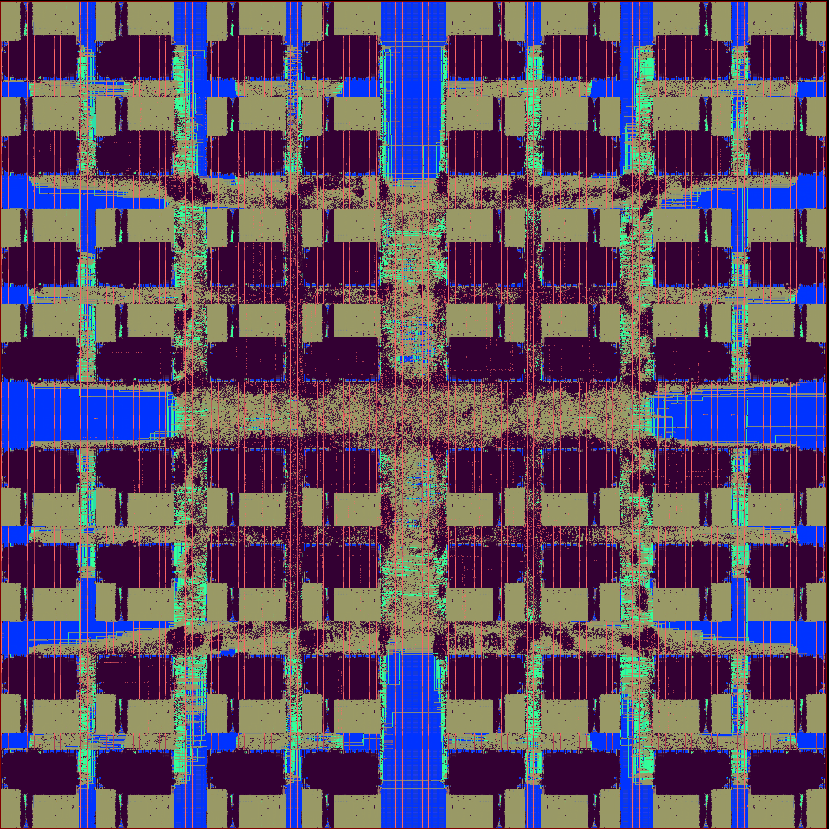}
    \subcaption{\topology{1}.}
    \label{fig:top1_be}
  \end{minipage}\hfill%
  \begin{minipage}[t]{0.50\linewidth}
    \centering
    \includegraphics[width=.9\linewidth]{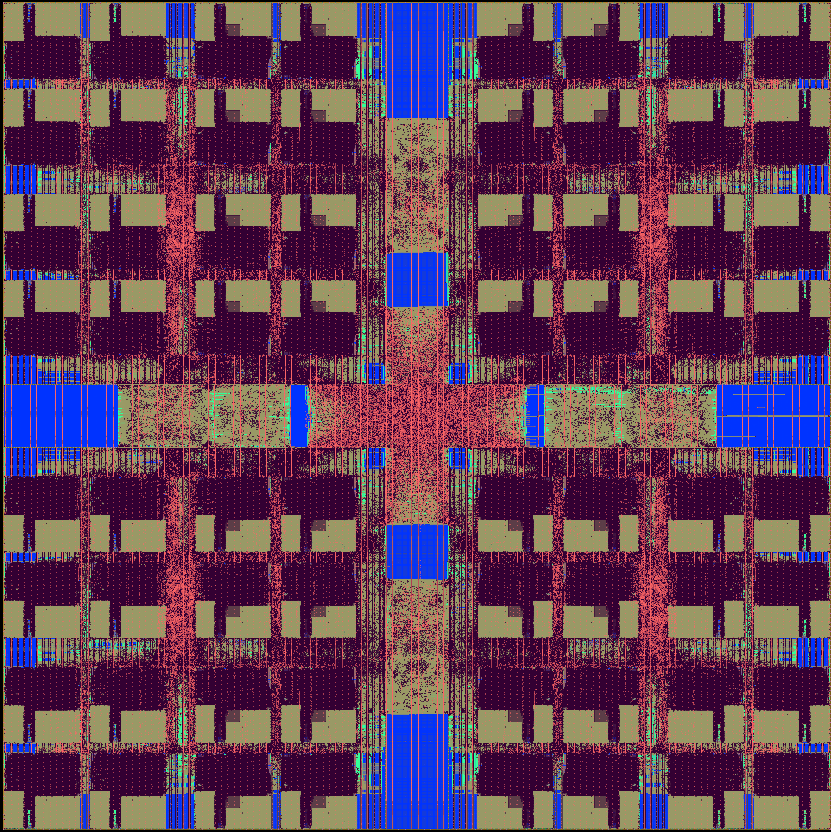}
    \subcaption{\topology{H}.}
    \label{fig:toph_be}
  \end{minipage}

  \caption{Placed and routed \topology{1} and \topology{H} \mempool{} clusters, implemented as \SI{4.6x4.6}{\milli\meter} macros.
    The dark blue regions are devoid of standard cells.
  }
  \label{fig:mempool_be}
\end{figure}

\Cref{fig:top1_be} shows the placed-and-routed \topology{1} macro.
With its \num{64x64} radix-\num{4} butterfly topology, the connection between any two remote tiles needs to cross the whole network, regardless of the physical distance between the tiles.
Therefore, all wiring and cells are drawn towards the center of the design, which is heavily congested.
\topology{4} is four times more congested than \topology{1}, which is enough to make it physically infeasible with reasonable clock rates.
The placed-and-routed \topology{H} macro can be see in \Cref{fig:toph_be}.
Similarly to \topology{1}, there is a high cell and wiring density at the center of the design, due to the connection between the two diagonally placed groups (\Cref{fig:arch_cluster}).
However, unlike \topology{1/4}, \topology{H} distributes the cells and the wiring throughout the cluster, through the use of the directional local group interconnects.

\topology{4} and \topology{H} achieve much better performance results (in terms of latency and throughput) than \topology{1}.
However, out of these two high-performance topologies, only \topology{H} is physically feasible.
The \topology{H} \mempool{} cluster runs at \SI{700}{\mega\hertz} at typical conditions (\SI{480}{\mega\hertz} at worst-case conditions).
The critical path of this design starts at the boundary of one local group, passes through the center of the cluster and another local group until reaching the \gls{ROB} of a Snitch core.
Wire propagation delay accounts for \SI{37}{\percent} of the timing of the critical path, and \num{27} out of the \num{36} gates in this path are either buffers or inverter pairs.

\subsection{Power analysis}
\label{sec:power-analysis}

In this section, we analyze the power and energy consumption of the \topology{H} \mempool{} cluster, while running the \emph{matmul} kernel at \SI{500}{\mega\hertz}, in typical operating conditions (TT/\SI{0.80}{\volt}\kern-.1em/\SI{25}{\celsius}).
Each tile consumes, on average, \SI{20.9}{\milli\watt}.
The main culprits are the instruction cache, at \SI{8.3}{\milli\watt} (\SI{39.5}{\percent} of the tile's total power consumption), the Snitch cores, at \SI{5.6}{\milli\watt} (\SI{26.6}{\percent}), and the \gls{SPM} banks, at \SI{2.6}{\milli\watt} (\SI{12.6}{\percent}).
The request and response interconnects only consume \SI{1.7}{\milli\watt}, less than \SI{10}{\percent} of the tile's total power consumption.
At the top level, \mempool{} consumes \SI{1.55}{\watt}, \SI{86}{\percent} of which being consumed within the tiles.

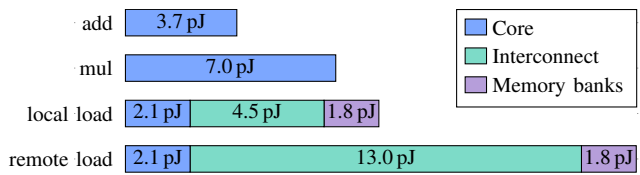
\begin{figure}[h]
  \centering
  \begin{tikzpicture}[/tikz/font=\footnotesize]
    \begin{axis}[
      xbar stacked,
      xmax=17,
      bar width = 1em,
      y axis line style = {draw=none},
      axis x line = none,
      tickwidth = 0pt,
      height = 3.75cm,
      nodes near coords,
      legend style={at={(1,1)}, anchor=north east, cells={anchor=west}},
      symbolic y coords = {remote load,local load,mul,add},
      ytick distance = 1,
      yticklabel style={align=right, xshift=4ex, font=\footnotesize},
      every pin/.style={font=\footnotesize},
      point meta = explicit symbolic]

      \addplot+[xbar] plot coordinates {(0,add) (0,mul) (0,local load) (0,remote load)};
      \addplot+[xbar, color=black, fill=color1!60] plot coordinates {(3.7,add) [\SI{3.7}{\pico\joule}] (6.986,mul) [\SI{7.0}{\pico\joule}] (2.148,local load) [\SI{2.1}{\pico\joule}] (2.148,remote load) [\SI{2.1}{\pico\joule}]};
      \addplot+[xbar, color=black, fill=color2!70] plot coordinates {(0,add) (0,mul) (4.45,local load) [\SI{4.5}{\pico\joule}] (13.0,remote load) [\SI{13.0}{\pico\joule}]};
      \addplot+[xbar, color=black, fill=color3!70] plot coordinates {(0,add) (0,mul) (1.82,local load) [\SI{1.8}{\pico\joule}] (1.82,remote load) [\SI{1.8}{\pico\joule}]};

      \legend{,Core, Interconnect, Memory banks}
    \end{axis}
  \end{tikzpicture}
  \caption{Breakdown of \topology{H}'s energy consumption per instruction.}
  \label{fig:energy}
\end{figure}

\Cref{fig:energy} summarizes the energy consumption per instruction of the \topology{H} tile, for different instructions.
Each local load uses \SI{8.4}{\pico\joule}.
About half of this energy consumption, \SI{4.5}{\pico\joule}, is spent at the local interconnect.
Remote loads use the global interconnect, which raises their energy consumption to \SI{16.9}{\pico\joule}.
In this case, the interconnects consume \SI{13.0}{\pico\joule}, or $\num{2.9}\times$ the energy consumed at the interconnects for a local load.

As a comparison with arithmetic instructions, a local load uses about as much energy as a complex instruction such as mul, or $2.3\times$ the energy consumed by a simple add.
Remote loads have the highest energy requirements, but even then that is only $4.5\times$ the energy of an add.
This result confirms that \mempool{} is a balanced design that is not severely interconnect-dominated.


\section{Conclusion}
\label{sec:conclusion}

In this paper, we presented \mempool{}, a \SI{32}{\bit} system with \num{256} ultra-small RV32IMA Snitch cores sharing \SI{1}{\mebi\byte} of L1 \gls{SPM}.
\mempool{} was implemented in \textsc{GlobalFoundries} 22FDX \gls{FDSOI} technology, achieving \SI{700}{\mega\hertz} in typical conditions.
\mempool{}'s architecture was driven by a physical-aware analysis of three different low-latency processor-to-L1-memory interconnect topologies.
We chose the one that leads to the best performance results in terms of throughput and average latency, while also being physically feasible.
In the absence of contention, all \gls{SPM} banks are accessible within \num{5} cycles.

We compared \mempool{}'s performance with a baseline system that has an idealized crossbar switch.
Our system achieves at least \SI{80}{\percent} of the baseline's performance on real-world signal processing benchmarks.
Our hybrid addressing scheme helps keep the memory requests in local banks accessible in one cycle, leading to performance gains up to \SI{20}{\percent} in such benchmarks.

Similarly to tile-based systems (which use 2D mesh \glspl{NOC}), this scheme provides low-latency access to a memory range.
However, our scheme has two advantages: (a) no aliasing, so that we can use this local range with more flexibility, and (b) much lower latency and higher bandwidth for all the global accesses, which enables us to run ``non-systolic'' algorithms effectively.
The addressing scheme is also highly efficient in terms of energy consumption since local memory requests consume only half of the energy required for remote accesses.


\bibliographystyle{IEEEtran} \bibliography{mempool}

\end{document}